%% file: Corona3rd.tex
\documentclass[aps,amsfonts,prd,nofootinbib]{article}

\title{Heterogeneity and Superspreading Effect on Herd Immunity}

\input{preamble.tex}
\usepackage{mwe}
\usepackage{float}
\usepackage[font=small]{caption}
\numberwithin{equation}{section}

\usepackage{fullpage}
\usepackage{authblk}

\author[1,2]{Yaron Oz}
\author[3]{Ittai Rubinstein}
\author[3]{Muli Safra}

\affil[1]{Raymond and Beverly Sackler School of Physics and Astronomy, Tel-Aviv University, Tel-Aviv 69978, Israel}
\affil[2]{School of Natural Sciences, Institute for Advanced Study, Princeton NJ, USA}
\affil[3]{Blavatnik School of Computer Science, Tel-Aviv University, Tel-Aviv 69978, Israel}

\date{\today}
\usepackage{theoremref}

\begin{document}
%
\maketitle

\begin{abstract}
We model and calculate the fraction of infected population necessary to reach herd immunity, taking into account the heterogeneity in infectiousness and susceptibility, as well as the correlation between those two parameters.
We show that these cause the effective reproduction number to decrease more rapidly, and consequently have a drastic effect on the estimate of the necessary percentage of the population that has to contract the disease for herd immunity to be reached.
We quantify the difference between the size of the infected population when the effective reproduction number decreases below 1 vs. the ultimate fraction of population that had contracted the disease.
This sheds light on an important distinction between herd immunity and the end of the disease and highlights the importance of limiting the spread of the disease even if we plan to naturally reach herd immunity.
We analyze the effect of various lock-down scenarios on the resulting final fraction of infected population.
We discuss implications to COVID-19 and other pandemics and compare our theoretical results to population-based simulations.
We consider the dependence of the disease spread on the architecture of
the infectiousness graph and analyze different graph architectures  
and the limitations of the graph models.  
\end{abstract}


%




\section{Introduction}


The COVID-19 pandemic has had a dramatic impact on the world in recent months, setting in motion a huge effort in diverse research disciplines in an attempt to understand the nature of the disease and the dynamics of the virus's spread.
A question of utmost importance in this context is how many infected individuals it takes to reach herd immunity, where by herd immunity one means that the virus is unable to find enough susceptible hosts to continue its spread and consequently the disease fades out. 

Herd immunity is typically expected to be reached when a large fraction of the population becomes immune to the disease and the effective reproduction number (which quantifies the spread) drops below one.
The standard estimate for the necessary fraction for herd immunity to be reached is about $60\%$ of the susceptible individuals.
This estimate, however, assumes a homogeneous structure of
the epidemic spread network, where both the infectiousness and susceptibility of individuals are assumed to be homogeneously distributed.

It is well recognized, however, that the epidemic spread network is
not homogeneous but rather heterogeneous, with distinct people being {\em infectious} (likely to infect others) and {\em susceptible} (likely to become infected themselves) to different degrees (for reviews see e.g. \cite{review1,review2}).
The class of individuals with very high secondary infection rates are referred to as superspreaders \cite{NatureCovid2}. An estimate for the COVID-19 pandemic \cite{Adi,dispersion} asserts
that between $5\%$ to $10\%$ of the infected individuals
cause $80\%$ of the secondary infections.

The reasons that different people are infectious and susceptible to different degrees may be, for instance, increased contact with others that can increase both parameters, or hygiene and better protective equipment that can decrease both. 
Thus, infectiousness and susceptibility are potentially highly correlated.
A correlation between infectiousness and susceptibility can significantly affect our estimate for the percentage of the population that must contract the disease for herd immunity to be reached. 

If more infectious people are also more susceptible, then our initial estimates of the basic reproduction number 
(the mean value of secondary infections caused by an infected individual)
will be biased leading us to believe that it is much larger than it really is---we are oversampling the infectiousness of more susceptible people.
Deviations in the susceptibility can lead to us seeing an early spike in the number of cases as the susceptible are infected, with a sudden drop later on as the disease spreads to less susceptible populations.
Furthermore, if more infectious people are also more susceptible, then they will also be infected and develop natural immunity much sooner.

Our aim in this work is to model and calculate the fraction of infected in the population that gives rise to
herd immunity while taking into account the heterogeneity in infectiousness and
susceptibility, their correlation and the superspreading effect.

It should be noted that contrary to some belief, even after herd-immunity is achieved, it may take a while for the disease to completely disappear and if there is a large fraction of infected population when the threshold is crossed, it may still infect many others during this decline. If, a strict counter-measure, such as a lock-down, is initiated near the same point as herd immunity is obtained, this "after-burn" effect can be greatly reduced. This misconception has led to a split in the related literature. Some results, such as \cite{Herd1}, model the beginning of herd immunity, while others, such as \cite{science,He} focus only on the total infected population (including the after-burn). This has led to two different and seemingly contradictory results about the size of the population that must be infected before the pandemic ends (if no vaccine is present).

In this paper we propose a mathematical model for the point where herd immunity is obtained and compare it both to population-based simulations and to the after-burn model.
We show that the herd-immunity models (which may predict far fewer total infected) are a good approximation for the end of a pandemic when the peak was kept in check, while the after-burn model (which can result in almost twice as many total infected) fits the case where no countermeasures are employed.
This result highlights the importance of taking precautions to limit the peak of the pandemic even if we plan to end the disease with a natural herd-immunity.

We will analyze the spread of the disease under the simplifying assumption that any two people can infect one another, that is the infection graph is complete.
We assign to each individual $a$ a susceptibility parameter $S(a)$ and an infectiousness parameter $I(a)$ drawn from some probability distributions. 
$S(a)$ and $I(a)$ quantify how likely $a$ is to be infected and
infect others, respectively.
The probability that $a$ will transmit the disease to $b$---should $a$ be infected---is $I(a) \times S(b)$ and this is the effective parameter in our analysis.
The product of $I$ and $S$ scales like the inverse of the total population size, where
as a convention we will set $S$ to scale like its inverse and $I$ to be independent of it.

We show that the temporal aspects of the infection are not relevant to the quantities that we analyze, namely the total number of infected individuals until herd immunity is reached and the after-burn effect.
This allows us to simplify our model by measuring the progress of the disease as a function of the number of individuals who contracted it, which
is the natural governing parameter when considering questions of the type: will the disease fade out as a function of the
fraction of the population that got infected.
We begin the evolution process of the disease at step $n=0$ (where by "step" we mean the number of infected individuals who have finished spreading the disease) with a certain number of infected individuals.
So long as the initial number of infected individuals is not very small, but also negligible compared to the size of the total susceptible population, it does not influence the analysis.
Our calculation is done by taking an expectation value over all possible scenarios of infection.

We model the system for any distribution of $I,S$, and present a general formula for the behavior of heterogeneous diseases.
We also consider the special case where each individual is infectious and susceptible to the same degree, that is, where the distributions from which $I$ and $S$ are drawn are highly correlated. 
We derive a simple analytical result when the infectiousness and susceptibility parameters are, in particular, chosen from a Gamma distribution with scaling and shape parameters $k$ and $\theta$, respectively---a distribution previously attributed to the infectiousness of COVID-2 \cite{NatureCovid2}.

It should be noted that such a framework, though similar to many recent models \cite{herd2, herd3, He, PRE_GCC}, is still far from perfect, as it overlooks many details of the spread of the disease, such as the specific contact graph, and we explore the limitations of such models in more detail. In particular, we provide two examples where the statistics of secondary infections seem identical, but in one $99.995\%$ of population is infected (when including the after-burn effect) and in the other only $o(1)$ fraction of the population is infected.





\subsection{Herd-Immunity}

Consider first the general case.
Define the average conditional infectiousness $\varphi(s)$:
\begin{equation}
\varphi(s) \defeq \Expect{S(a)=s} {I(a)} \ ,
\label{phi}
\end{equation}
and the normalized susceptibility distribution at the step $n$ in the evolution of the disease $\rho(s,n)$:
\begin{equation}
\rho(s, n) \defeq \cProb{}{a \;\mbox{is healthy after $n$ steps}}{(S(a)=s)}  \ .
\label{rhon}
\end{equation}

Denote the susceptibility distribution at the beginning by $\rho(s) = \rho(s,0)$.
The average conditional infectiousness (\ref{phi}) is independent of the number of infected individuals while the susceptibility
distribution does depend on it---individuals with higher susceptibility are more likely to be chosen first, hence their rate decreases as the process progresses.

$\rho(s)$ and $\varphi(s)$ are the fundamental
quantities that define our model. 
Our derivations are general and the results can be applied for any $\rho(s)$ and $\varphi(s)$. 
In the particular case of the Gamma distribution, commonly used to describe infectious diseases, we can derive a precise
analytical formula (Claim II).  
The canonical examples of infectious diseases are $\varphi(s)\propto s$ (infectiousness and susceptibility are fully correlated) and $\varphi(s) = constant$ (infectiousness and susceptibility are not correlated) and we compare the effects of these scenarios.

We will prove the following general claim about the 
fraction of the population necessary to reach herd immunity:
\vskip0.1cm
{\bf Claim I} (General Case): 
{\it for any $\delta$ when}
\begin{equation}
    1 - \int \rho(\sigma) exp(-\delta \sigma) d\sigma
    \label{f}
    \end{equation}
{\it     fraction of the population is infected, the effective reproduction number will be reduced by a factor of}
\begin{equation}
    \frac{\int \varphi(\sigma) \rho(\sigma) \sigma \exp\left(-\delta \sigma\right) d\sigma}{\int \varphi(\sigma) \rho(\sigma) \sigma d\sigma} \ .
    \label{fraction}
\end{equation} 
The threshold for herd immunity is when the value of the effective reproduction number is 1.

Consider next the particular case of the Gamma distribution with shape and scale
parameters $k$ and $\theta$, respectively and $\varphi(s)\propto s$.
We will prove the following claim:
\vskip 0.1cm
{\bf Claim II} (Gamma distribution): 
    {\it Under the above assumptions, herd immunity will be reached when}
    \begin{equation}
    \varepsilon=1-{R}^{\frac{-k}{(k+2)}}
    \label{eps}
\end{equation}    
{\it    fraction of the population is infected. $R$ is the reproduction number at the beginning of the disease spread and $k$ is the shape (spread) parameter of the Gamma distribution.}
\vskip 0.1cm

Substituting the estimates for COVID-19: $R \approx 3$  and $k \approx 0.1$ \cite{dispersion,us}, we get $\varepsilon\approx5\%$.
This result is far more optimistic than the recent estimate \cite{science} that requires about $40\%$ of the population to contract the disease before herd immunity is achieved. 
The Gamma distribution has been used in order to model the data of COVID-2
with  $k=0.19$ \cite{NatureCovid2} and in this case we get $\varepsilon\approx9\%$.

\subsection{The After-Burn Effect}

One of the first things that one notices when running population-based simulations is that even after the basic reproduction number goes below one, it can still take a long time for the disease to disappear. We call this phenomenon the "after-burn effect".

Previous works in this field such as \cite{He,PRE_GCC} quantify the effects of heterogeneity on the total number of infected including the after-burn by calculating the size of the giant--connected-component (GCC) of the random graph. As explained above, this is different from the herd-immunity model that neglects this effect resulting in a significantly more optimistic projection for the end of the pandemic.

In \cite{PRE_GCC}, the authors analyze random undirected graphs where the degree of the nodes can be drawn from arbitrary distributions, while in \cite{He} the authors use the results of \cite{PRE_GCC} to analyze the case where the degrees are chosen according to a negative binomial distribution and compare the results to various pandemics.
In our model, we consider a directed graph where the in-degree is correlated to the out-degree but not necessarily equal to it.
We show how the results of \cite{PRE_GCC} can be extended to this case, and compare them to population based simulations.

\paragraph{Herd Immunity versus GCC:}
The comparison between the herd-immunity model and the GCC one is not as trivial as it may seem. The herd-immunity model, despite neglecting an important aspect of the pandemic, may be more realistic in some cases, since in order to reach a strong after-burn effect, a large number of individuals must be infected when we reach herd immunity and this does not happen when strict counter-measures are employed at peaks of the disease (as is usually the case), or if a final lockdown begins just as the basic reproduction number goes below one.

In other words, we can think of the herd-immunity model as being a model for a society that either keeps the size of the total infected population in check or has a final lock-down just as herd-immunity sets in and the after-burn model as a society which takes no steps to limit the spread of the pandemic. In our simulations we considered all three cases (constant countermeasures, countermeasures after herd immunity and no countermeasures) and show that they end up close to the predicted outcomes.
Thus, studying and comparing these two models is crucial for our understanding of heterogeneous pandemics as they show us the expected outcomes of pandemics with and without controlled countermeasures.

\subsection{Organization}
\vskip 0.5cm
The paper is organized as follows. In section II we will define precisely the 
effective reproduction number that we will use in our analysis. In section III we will study the dynamics of the disease spread and prove claim I. In section IV we will prove claim II and study numerically claim I for various types of infection and susceptibility distributions. In Section V we will adapt the GCC based analysis from \cite{PRE_GCC} to our case and consider compare the GGC calculations and our
population based simulations. We will discuss the time independence of the
total number of infected individuals until herd immunity is reached and the after-burn effect.
We will consider the effects of various countermeasures and the correlation between susceptibility
and infectiousness.
In Section VI we will showcase the limitations of models that overlook the details of the graph architecture. Section VII is devoted to a discussion and outlook.

\section{Effective Reproduction Number}

Consider the effective value of the reproduction number at step
$n$ in the evolution of the pandemic which we will  denote by $R(n)$,
$R=R(0)$.
We define it as the expectation value of secondary infections
conditional on the individuals that have been infected.
Denote by $\Lambda_n$ the distribution over the $n$th individual to be infected (namely, linear in susceptibility at that stage) --- we thus have:
\begin{equation}
R (n) \defeq \Expect{a\sim\Lambda_n}{\mbox{number of people $a$ infects}} \ .
\label{rb}
\end{equation}

Mathematically we have:
\begin{equation}
    \begin{aligned}
R(n)
    &=  \Expect{a\sim\Lambda_n} {I(a) \sum_{b\neq a} S(b)} =\\
    &= \sum_a \frac{S(a)}{\sum_{a'} S(a')} I(a) \sum_{b\neq a} S(b) =\\
    & \simeq \sum_a S(a) I(a) \ , 
    \end{aligned}
    \label{rbm}
\end{equation}
where the summation over $a$ in the last two lines is on the healthy individuals at step $n-1$ and in the approximation at the last line we added the infected individual $a$ to the summation.

In the final step we used the approximation:
\begin{equation}
\frac{\sum_{b\neq a} S(b)}{\sum_{a'} S(a')} = 1 - \frac{S(a)}{\sum_{a'} S(a')} = 1 - O\left(\frac{\max_a (S(a))}{\sum_{a'} S(a')}\right) \ ,
\end{equation}
which means that our approximation is valid so long as no individual infects a significant fraction of the population. In particular, when the infectiousness and susceptibility are taken from a Gamma distribution, this error term is bounded by 
\begin{equation}
\frac{\max_a (S(a))}{\sum_{a'} S(a')} = O\left(\frac{\log N}{N}\right) \ .
\end{equation}

The RHS of equation (\ref{rbm}) may look confusing as it seems to resemble the homogeneous case:
\begin{equation}
R_0 = \frac{1}{N} \sum_{a,b} I(a) S(b) \ ,
\end{equation}
where $I(a) S(b)$ has the physical interpretation that $a$ infects $b$. Thus, it looks as if in (\ref{rbm}) $a$ infects itself.
However, the meaning of $S(a) I(a)$ is the probability of $a$ being infected times the expected number of people $a$ will infect if infected, thus giving us the expected number of people $a$ will infect.
The fact that an expression as simple as this turns out to be the correct way to combine the correlation between individual infectiousness and susceptibility allows us to greatly simplify the rest of our derivation.

Using (\ref{phi}) and (\ref{rhon}) we get:
\begin{equation}
 R(n) = N(n) \int \varphi(s) \rho(s,n) s ds \ .   
 \label{rne}
\end{equation}
We denote the susceptible population at the beginning by $N=N(0)$.

In the next section we will study the dynamics of the spread of the disease and how the effective reproduction number $R(n)$ depends on the number of infected individuals.

\section{Dynamics of the Spread of Diseases}

The process starts at step $n=0$ with a few infected individuals.
At each step that someone is infected, the probability that $a$ was infected is proportional to $S(a)$.
Thus,
\begin{equation}
    \begin{aligned}
&\Prob{}{a\mbox{ is healthy at step } n } =\\
        &\;\;\;\;= \left(1 - \frac{S(a)}{\sum_{b\colon b\mbox{ is healthy at step } n-1} S(b)}\right) \cdot \\
        &\;\;\;\;\cdot \Prob{}{a\mbox{ is healthy at step } n-1} \ .
    \end{aligned}
\end{equation}
We therefore get:
\begin{equation}
    \begin{aligned}
\log(&\Pr{\left( a\mbox{ is healthy at step }n\right)}) -\\
&-\log\left(\Pr{\left( a\mbox{ is healthy at step }n-1\right)}\right)= \\
    &=-\frac{S(a)}{N(n-1)\Expect {b\sim\Lambda_{n-1}} {S(b)}} \\
    &- O\left(\left(\frac{S(a)}{\sum_{b\mid b\mbox{ is healthy at step }n-1} S(b)}\right)^2\right)\\
    &= -\alpha(n) S(a) - O\left(\left(\frac{\max_b {S(b)}}{N \mathbb{E}_b S(b)}\right)^2\right) \ ,
    \end{aligned}
\end{equation}
Where
\begin{equation}
\alpha(n) = \frac{1}{\sum_{b\mid b\mbox{ is healthy at step }n-1} S(b)} \ .
\end{equation}
Since at each step someone is infected, there can be at most $N$ steps:
\begin{equation}
    \begin{aligned}
\log( & \Prob {} {\mbox{a is healthy at step $n$}} ) = \\
    &= -\left(\sum_{\tau\leq n} \alpha(\tau)\right) S(a) - O\left(\left(\frac{\max_b {S(b)}}{N \mathbb{E}_b S(b)}\right)^{2} n\right) =\\
    &= -\left(\sum_{\tau\leq n} \alpha(\tau)\right) S(a) - O\left(\left(\frac{\max_b {S(b)}}{\mathbb{E}_b S(b)}\right)^{2}  \frac{n}{N^2}\right) \ .
    \end{aligned}
\end{equation}

Therefore, as long as $\frac{\max_b {S(b)}}{\mathbb{E}_b S(b)} \ll \sqrt{N}$ we have:
\begin{equation}
    \begin{aligned}
\log(&\Prob{}{a\mbox{ is healthy at step }n}) \approx \\
    &\approx -\left(\sum_{\tau\leq n} \alpha(\tau)\right) S(a) \ .
    \end{aligned}
\end{equation}

Denote $\delta(n) = \sum_{\tau\leq n} \alpha(\tau)$, we get a relation between the susceptibility distributions at steps $n$ and zero:
\begin{equation}
    \begin{aligned}
\frac{\rho(s,n)}{\rho(s,0)} = \frac{exp(-\delta(n) s)}{ \int \rho(\sigma, 0)exp(-\delta(n) \sigma)  d\sigma} \ ,
    \end{aligned}
    \label{rho}
\end{equation}
between 
the susceptible populations: 
\begin{equation}
\frac{N(n)}{N(0)} = \left(\int \rho(s, 0) exp(-\delta(n) s) ds\right)  \ ,
\label{N}
\end{equation}
and between the effective reproduction numbers (\ref{rne}):
\begin{equation}
\frac{R(n)}{R(0)} =  \frac{ \int \varphi(s) \rho(s, 0)exp(-\delta(n) s) s ds}
{\int \varphi(s) \rho(s, 0) s ds}  \ . 
\label{R}
\end{equation}
This proves Claim I with (\ref{f}) being $1-\frac{N(n)}{N(0)}$.

It might be surprising that equation (\ref{rho}) does not depend upon the infectiousness distribution, but this should be expected as it merely encodes the order in which different individuals are infected (giving the probability that an individual with susceptibility $s$ is not among the first $n$ infected) and this is determined solely by susceptibility.

We can derive the condition for reaching herd immunity at step $n_{herd}$ by:
\begin{equation}
    \begin{aligned}
1 &= R (n_{herd}) = N(n_{herd}) \int \varphi(\sigma) \rho(\sigma, n_{herd}) \sigma d\sigma =\\
    &= N(0) \int \varphi(\sigma) \rho(\sigma, 0) \exp(-\delta(n_{herd}) \sigma)   \sigma d\sigma \ ,
    \end{aligned}
\end{equation}
where $N (n_{herd})$ is related to $N(n=0)$ by (\ref{N}).

\section{Herd-immunity}
Consider first the particular case where the infectiousness and
susceptibility
$I\sim S$ and both are drawn from a Gamma distribution:
\begin{equation}
 \rho_{k, \theta} (s,0) =    
 \frac{1}{\Gamma(k)\theta^k} {s}^{k-1} \exp\left(-\frac{s}{\theta}\right) \ ,  
\end{equation}
where $k$ and $\theta$ are shape and scale parameters of the distribution at $n=0$.
Using (\ref{rho}) we get:
\begin{equation}
\rho_{k, \theta} (s,n) = \rho_{k, \theta(n)} (s, 0),~~~~\theta(n) = \frac{\theta}{1 + \theta\delta(n)} \ .
 \label{eqn:miracle}
\end{equation}
Thus, while the shape of the distribution does not 
change during the evolution of the disease its scale does.

We denote 
  $\beta(n) = \frac{\theta(n)}{\theta}$,  
then using (\ref{N}) and (\ref{R}) we get:
\begin{equation}
\frac{N(n)}{N(0)}  = \beta(n)^k,~~~~~\frac{R(n)}{R(0)} =  {\left(\frac{N(n)}{N(0)}\right)} ^ {\frac{k+2}{k}} \ ,
\end{equation}
where 
\begin{equation}
    \begin{aligned}
R(n) 
= N(n) \int \rho(s,n) {s}^{2} ds 
    =\left(k^2 + k\right) \theta^2(n) N(n) \ .
    \end{aligned}
\end{equation}

Herd immunity is reached when $R(n)$ drops below one and this happens when
\begin{equation}
\frac{N(n)}{N(0)} = R^{\frac{-k}{k+2}} \ ,
\end{equation}
thus, proving claim II.

Calculating the fraction of the population $\varepsilon$ leading to herd immunity, following claim I, can be carried out analytically for the case of a Gamma distribution, while for general distributions it has to be performed numerically.
In figures 1 and 2 we plot $\varepsilon$ as a function
of the coefficient of variation, i.e. the ratio of the standard deviation and the mean,  
${\rm Cv}=\frac{\sigma}{\mu}$.
We plot the results for three distributions: Gamma, Folded Normal (Truncated Gaussian) and Power Law when $R$ equals three and six, respectively. 

We set $\varphi(s)=s$, hence $R$ is the second moment of the distribution.
We can see that as ${\rm Cv}$ approaches $0$, the distributions behave similarly. However, for larger values, the behavior of the system depends upon the distribution with the power law distribution approaching $\varepsilon = 0$ much faster.
The coefficient of variation of the truncated Gaussian distribution converges to $\sqrt{\frac{\pi - 2}{2}} \approx 0.755$, that is its value for the
 Half-Normal distribution to which the distribution converges as 
$\frac{\mu}{\sigma}\rightarrow 0^+$.

In general, we see that the higher the variance of the 
infectiousness and
susceptibility the lower the fraction of the population that needs to be infected in order to reach herd immunity.

\begin{figure}[H]
\centering
\includegraphics[width=0.7\columnwidth]{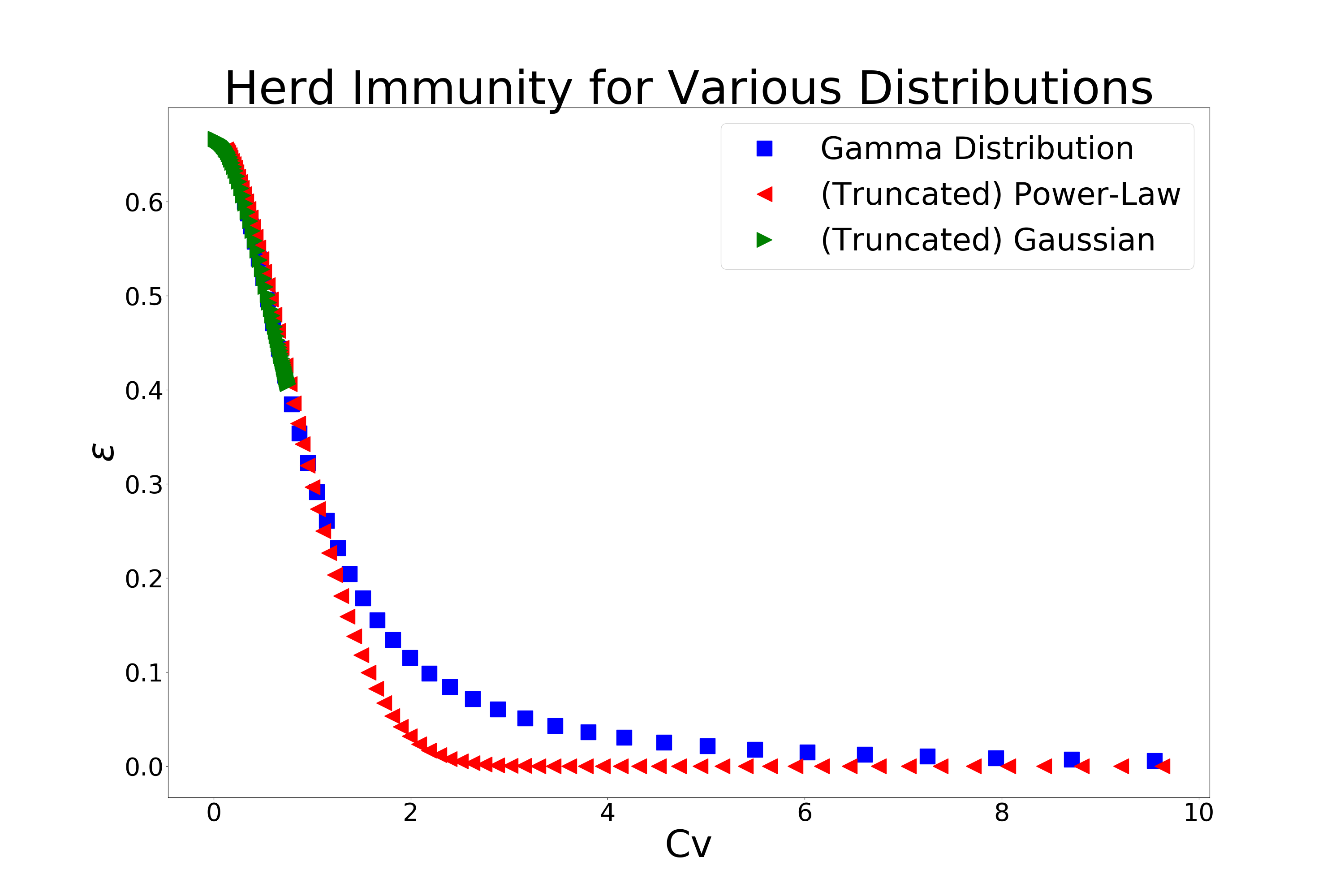} 
\caption{The percentage of infected in the population necessary for herd immunity to occur as a function of the coefficient of variation for $R=3$.
The results are shown for three distributions: Gamma, Folded Normal and Power Law. 
The higher the variance of the 
infectiousness and
susceptibility the lower the fraction of the population necessary. 
}
\end{figure}

\begin{figure}[H]
\centering
\includegraphics[width=0.7\columnwidth]{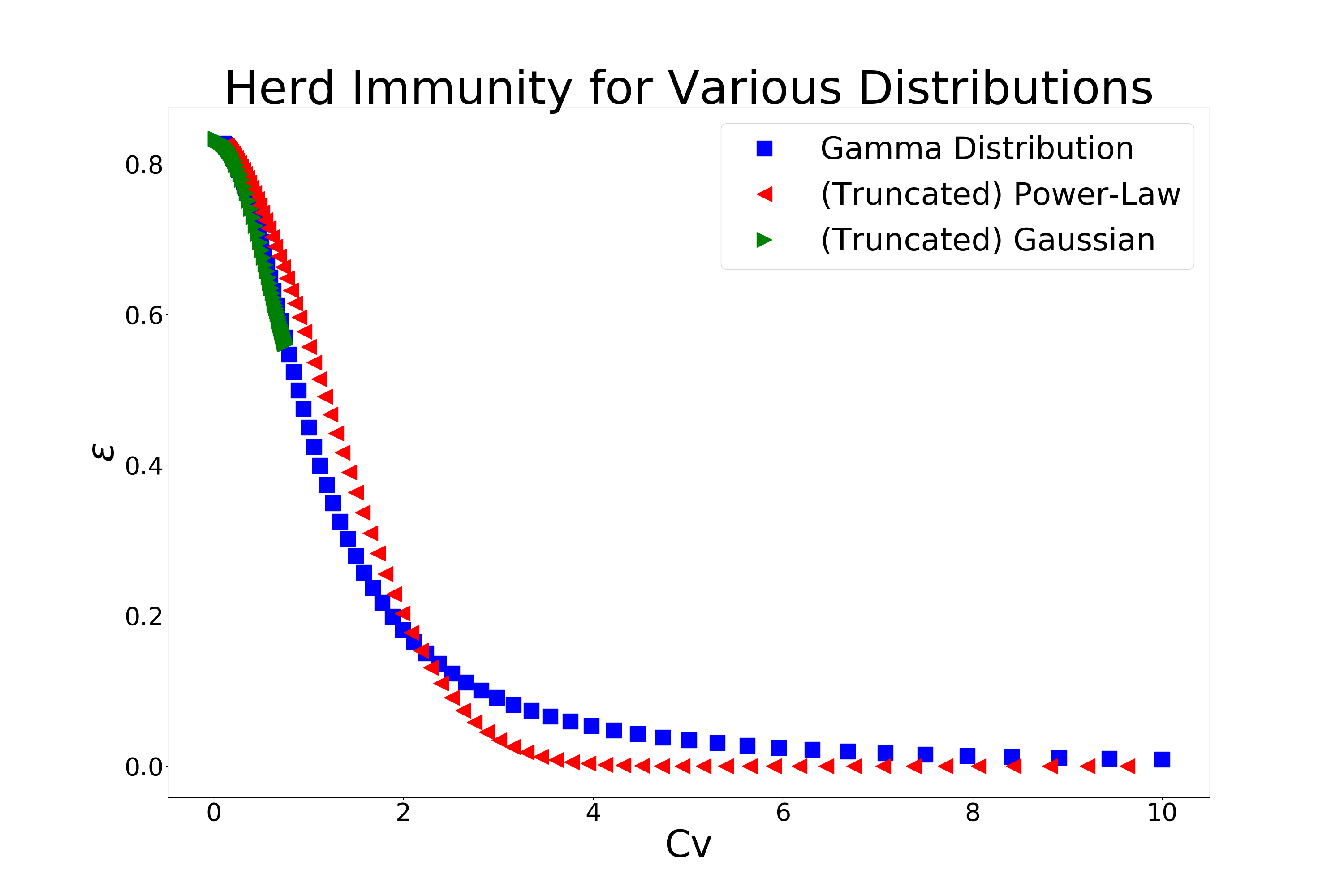} 
\caption{The percentage of infected in the population necessary for herd immunity to occur
as a function of the coefficient of variation for $R=6$.
The results are shown for three distributions: Gamma, Folded Normal and Power Law. 
In comparison to figure 1 we see, as expected, that the required 
fraction is higher for a given coefficient of variation.
}
\end{figure}

\section{Population-based Simulations}
This section involves several population-based simulations, which will allow us to validate our analytical results, study the after-burn effect and analyze various counter measures that can be taken and their effectiveness. 

\subsection{Methods} \label{sec:Methods}

Each of our population-based simulations considers a set of individuals. 
Each individual has susceptibility and infectiousness parameters randomly assigned according to prescribed probability distribution;
they can be either "Healthy", "Infected" or "Recovered" at every point in time.
We initialize almost all of the individuals to "Healthy", and only a small percentage to "Infected".
We observed that when the number of initially-infected is not extremely small or very large it does not significantly affect the spread of the disease, so we set it to $1000$ individuals.
We then maintain a FIFO queue of the infected population. In each step we take the "oldest" infected individual $A$ and for each healthy individual $B$ we set $B$ to be "Infected" with probability $A.infectiousness\times B.susceptibility$, and then set $A$ to be "Recovered".
We allow several types of countermeasures to be employed, reducing the infection probability where appropriate.
In figure \ref{fig:pop_sim_temporal} we show several example runs of our simulation with different countermeasures.

\subsection{Time Independence}
Some previous models (such as \cite{Herd1}) take into account the time delay between exposure and infection, as well as the time it takes for an infected individual to recover.
While these effects are crucial for our understanding of the time-dependence of a disease and for answering questions such as "what will be the peak number of infected?" or "how long should the lockdown be?", we claim that they do not affect the total number of infected individuals until herd immunity is reached or the after-burn. Showing that these parameters of the model are not important allows us to simplify our model and obtain more analytical results.

To see this, as in \cite{He}, we consider a graph of the population where an edge exists between two individuals (nodes in the graph) $A$ and $B$ if, should $A$ become infected while $B$ is healthy $A$ will infect $B$. In standard models, each time an individual is infected, the existence or nonexistence of each of its outgoing edges is decided by some random process. 
However, since this process is independent of the current state of the system, we could have determined these edges in advance, at which point the total number of infected comes down to a question of the size of the GCC of the graph.
Thus, the  
time delay between exposure and infection, as well as the time it takes for an infected individual to recover, do not affect the fraction of the population that suffice to reach herd immunity or the end of the disease following
the after-burn effect.

\subsection{Modeling the After-Burn in the Homogeneous Setting}

Thinking in terms of the GCC can help us clarify the difference between herd immunity and the total infected population at the end of the disease, by analyzing the homogeneous disease spread distribution scenario. In the homogeneous case the epidemic graph is a directed random graph $D(N,p)$ with $N$ vertices (population size) 
and probability for an edge between two nodes (infection) $p=\frac{R_0}{N}$. 
$R_0$ denotes the basic reproduction number.
Such random graphs have been studied by many Mathematicians and Computer Scientists
\cite{Alon}, and it was shown \cite{Karp} that number of vertices that can be reached from any one vertex in such a graph is almost always approximately $\Theta N$, where $N$ is the size of the population and $\Theta$ is the root of $1 - x - e ^ {-R_0 x} = 0$ in the segment $\left(0,1\right)$.

Similarly, even when considering the relationship between the total number of infected and the current reproduction rate $R(N)$, it is clear that, under our model's assumption that each individual's probability to be infected is dependent only on their susceptibility, the temporal aspect is irrelevant since the distribution of who is or isn't healthy at any given point depends only on the number of infected so far.

\subsection{Modeling the After-Burn in the Heterogeneous Setting} \label{sec:GCC_theory}

In \cite{PRE_GCC}, the authors present a method of computing the size of the GCC of a random undirected graph with arbitrary degree distribution.
Their derivation is based upon the careful analysis of several generating functions.
In particular, they define 
\begin{equation}
G_0 (x) = \sum_k p_k x^k \ ,
\end{equation}
to be the generating function of the distribution of the degree of a random vertex in their graph, and $G_1 (x)$ to be the generating function of the distribution of the degree of a vertex chosen by advancing through a random edge. In the case of undirected graphs, this distribution is obtained by weighting each vertex by its degree and removing the $1$ edge used to enter it and as a result, for undirected graphs (see equations (8), (9) of \cite{PRE_GCC})
\begin{equation}
G_1 (x) = \frac{\sum_k k p_k x^{k-1}}{\sum_k k p_k} = \frac{G_0 ' (x)}{G_0 ' (1)} \ .
\end{equation}
They then define two more generating function: $H_0 (x)$ and $H_1 (x)$ which are the generating functions of distributions of the sizes of the components reached from a random vertex or edge respectively.

The main difference of their analysis and ours the use of $H$. In \cite{PRE_GCC},
the authors use the fact that when there is a giant component, its size will be $1 - H_0 (1)$ since $H_0 (1)$ is the probability that a component will have a finite size. In other words, they analyze the probability that the part of the graph {\em reached from} a random vertex is large, while we want to analyze the probability that the part of the graph that {\em reaches} a random vertex is large. Since our graphs are directed these questions are not equivalent.

Note, however, that we can still use their analysis on the reversed infection graph (that is we will reverse all the edges in our graph and then apply the $H$ analysis to find the probability that a random vertex reaches a large portion of the graph). So when using this analysis for directed graphs, we define $G_0 (x)$ to be the generating function of the distribution of {\em in-degrees} (of the original graph), and $G_1 (x)$ to be the distribution of in-degrees weighted by the out-degrees.
Furthermore, since our graph is directed and it is extremely unlikely that both an edge and its reverse will exist, we do not need to remove $1$ from out-degree, giving us 
\begin{equation}
G_1 (x) = \frac{\sum_k k p_k x^{k}}{\sum_k k p_k} = x \frac{G_0 ' (x)}{G_0 ' (1)} \ .
\end{equation}

\begin{figure}[H]
\centering

\includegraphics[width=.7\columnwidth]{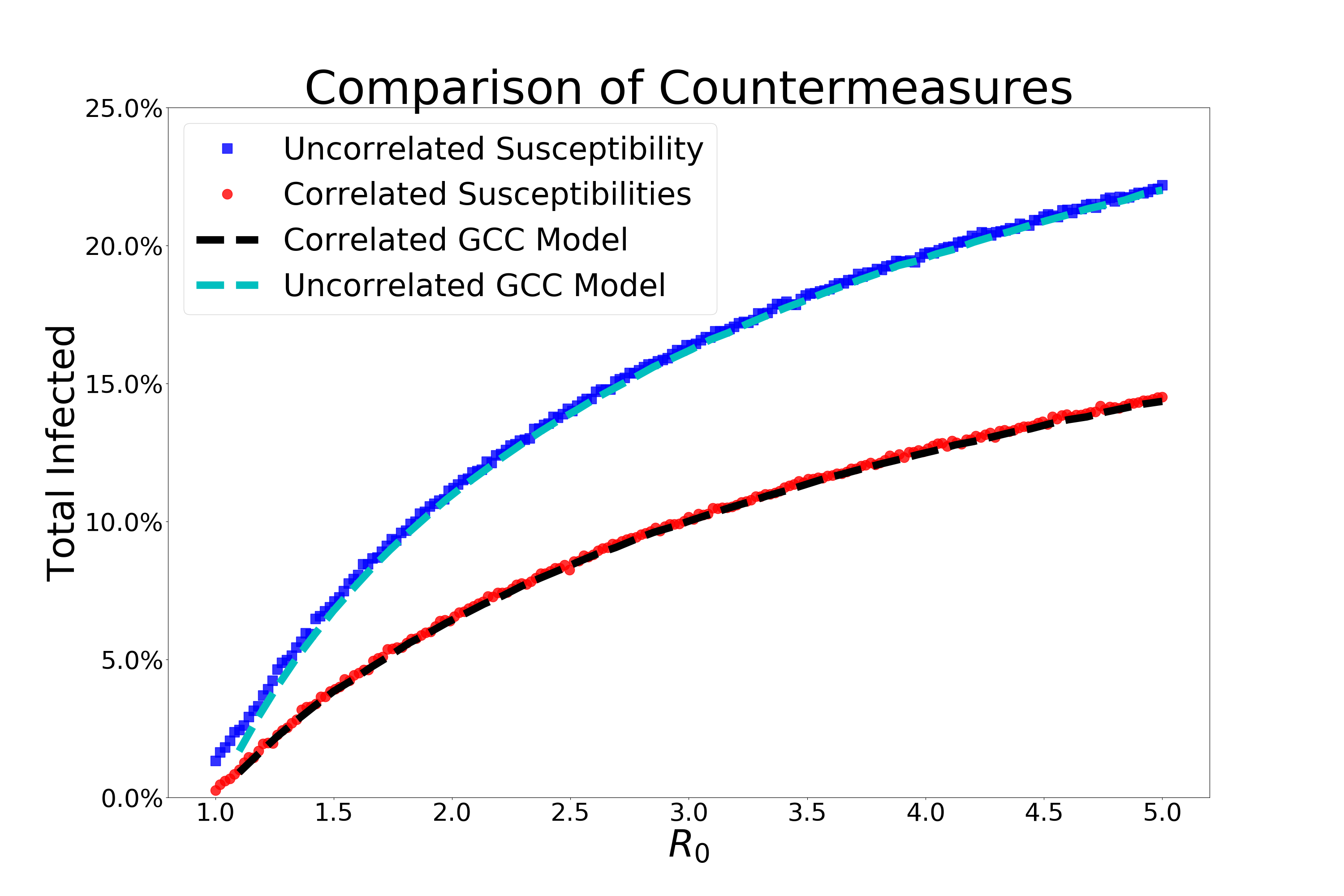} 
\caption{
    We compare our theoretical analysis of the size of the GCC of general directed graphs (shown in section \ref{sec:GCC_theory}) to an independent population-based simulation (defined in section \ref{sec:Methods}). In each simulation we set the size of the population to $N=1000000$, and derive the susceptibility and infectiousness of each individual either in a correlated manner ($\varphi(s) = s$), or in an uncorrelated manner ($\varphi(s) = 1$), setting the distribution of the susceptibilities to a Gamma distribution with shape $k=0.1$ (and varying the scale so that the desired reproductive number $R_0$ is obtained). As we can see, in both cases, the total infected population behaves as expected.
    }
    \vskip-1em
\label{fig:correlated_uncorrelated}
\end{figure}

Therefore, in our analysis we set 
\begin{equation}
G_0 (x) = \int G_{Poisson(\Phi s)} (x) \rho(s) ds = \int \exp{\left(\Phi s (1 - x)\right)} \rho(s) ds \ ,
\end{equation}
and 
\begin{equation}
G_1 (x) = N\frac{ \int \exp{\left(\Phi s (1 - x)\right)} \rho(s) \varphi(s) ds}{\Phi}
\end{equation}
where 
\begin{equation}
\Phi = N\int \rho(s) \varphi(s) ds \ .
\end{equation}
The rest of the analysis stays the same, and the fraction of the population infected by the disease is $1 - G_0 (u)$ where $u$ is the solution to the equation $u = G_1 (u)$.
We test this analysis using our population-based simulations
when infectiousness and susceptibility are correlated: 
\begin{equation}
\varphi(s) = s \ ,
\end{equation}
and when they are not correlated:
\begin{equation}
\varphi(s) = \varphi(0) \ .
\end{equation}
We compare the results in figure \ref{fig:correlated_uncorrelated}.

\subsection{Results}

\begin{figure}[H]
\centering

\includegraphics[width=.7\columnwidth]{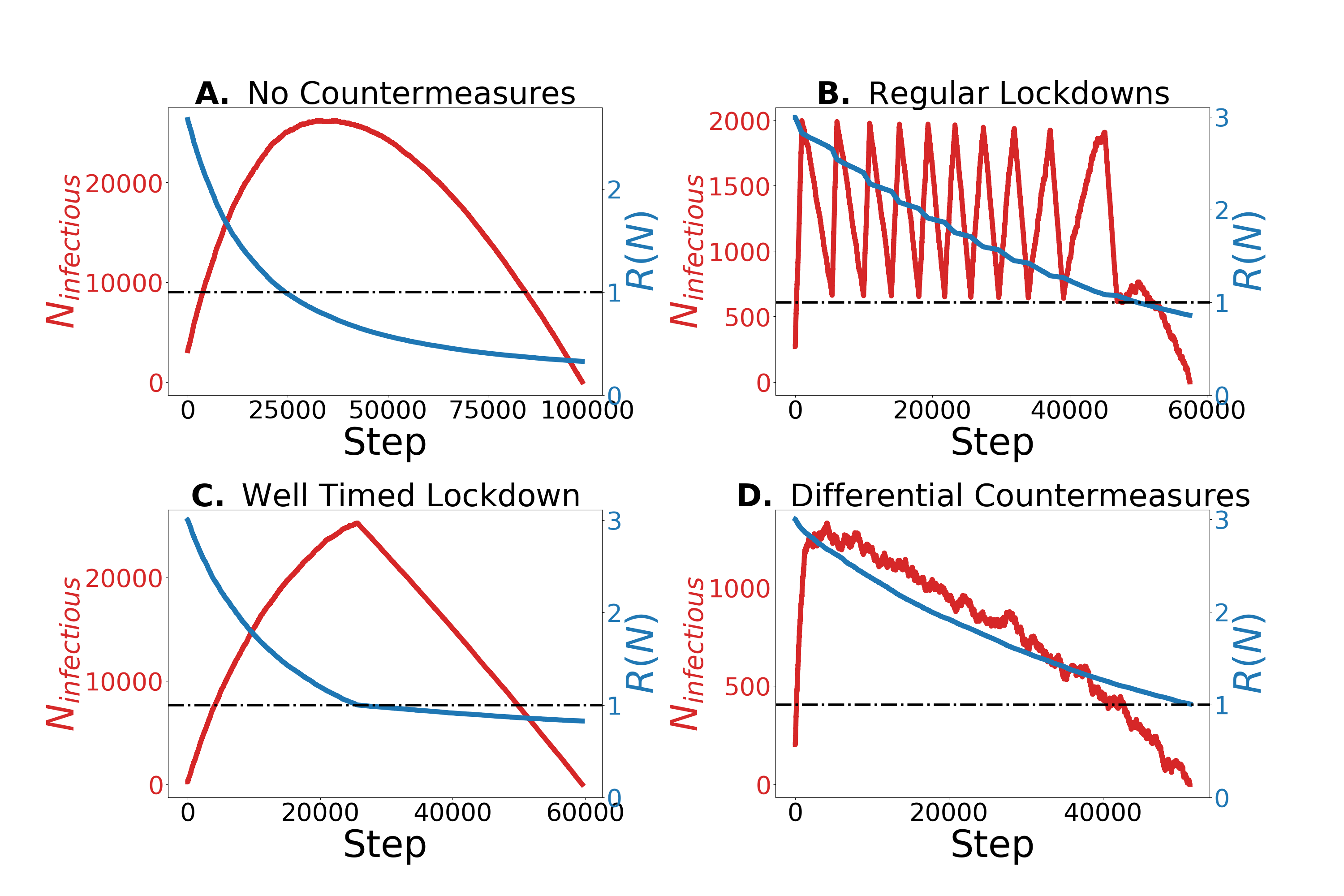} 
\caption{
    We simulated four types of populations each of size $N=1000000$, with $k=0.1$ (as estimated for COVID-19 \cite{dispersion, us}) and a reproductive rate of $R_0=3$. We plotted the progress of their diseases where the $x$ axis corresponds to the steps of the simulator (the number of recovered).
    {\bf A} 
    The first population takes no steps to prevent the spread of the disease. As we can see in this case, the disease tends to continue spreading long after herd-immunity is reached. 
    {\bf B}
    The second population goes into lock-down whenever the percentage of the population currently infectious crosses a set threshold ($0.2\%$ in our simulations) and goes out of lock-down when it goes under another set threshold ($0.05\%$ in our simulations). As a result, the size of the infected in this model oscillates between these two thresholds until a sufficient level of herd immunity is reached, causing the disease to fade away shortly after herd immunity.
    {\bf C}
    The third population goes into a single lock-down when the reproductive number goes below $1$. As we can see, their infected population starts off growing in a similar manner to the first population before entering a lockdown until the disease fades away shortly after herd immunity.
    {\bf D}
    The fourth population constantly takes preventative measures to keep their total infected population below the threshold of entering a lock-down (taking more preventative measures the closer they are to it). As a result, the size of the infectious population maintains a slow downward slope as the reproductive number decreases until the infection ends almost immediately upon reaching herd immunity.
    }
    \vskip-1em
\label{fig:pop_sim_temporal}
\end{figure}

\begin{figure}[H]
\centering

\includegraphics[width=.7\columnwidth]{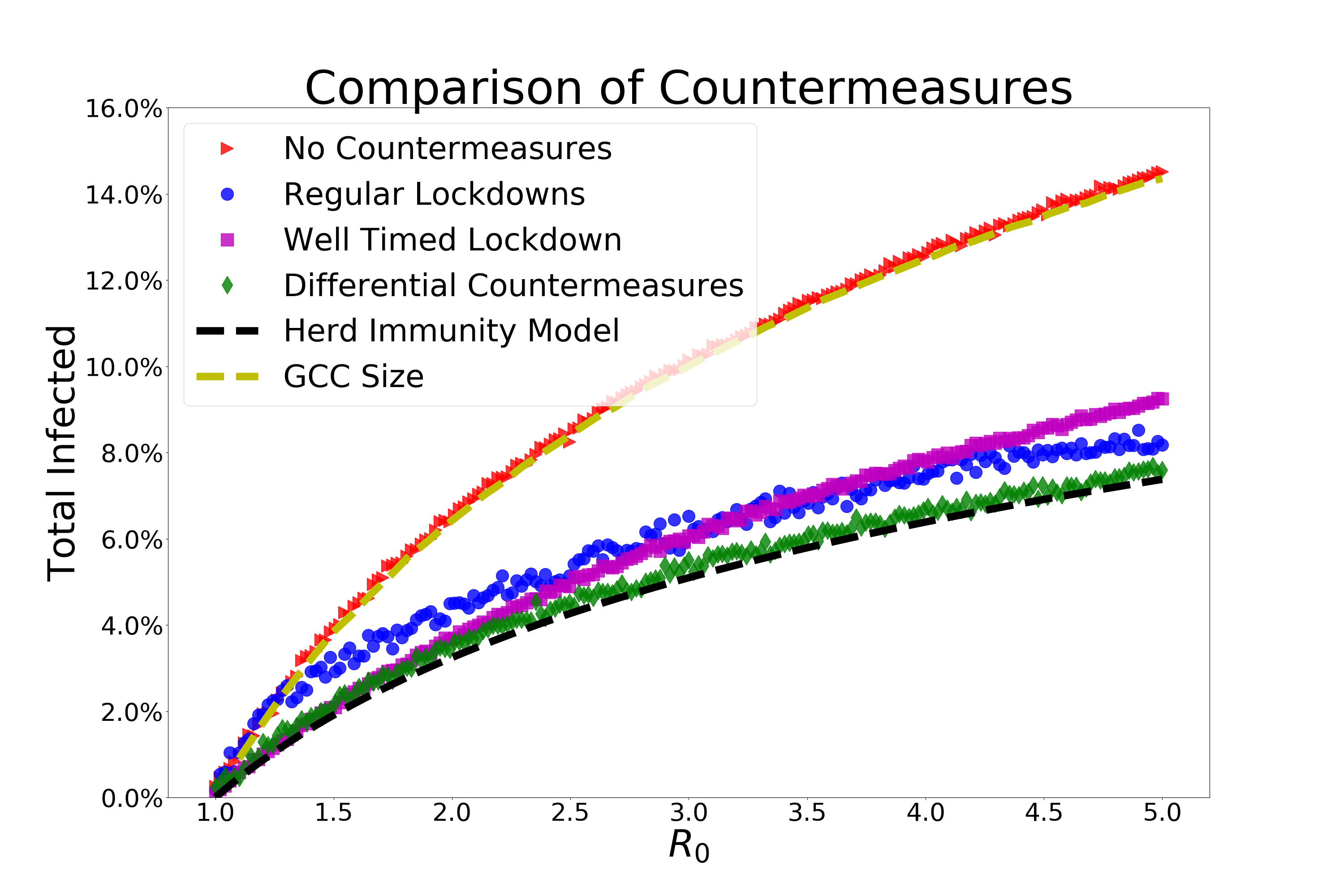} 
\caption{
    We compare our results to previous works in this field and to several population-based simulations. Using population-based simulations as in figure \ref{fig:pop_sim_temporal} but with a reproductive number $R_0$ running from $1$ to $5$, we plotted the total infected population as a function of $R_0$. As expected, we see that the population that takes no countermeasures has a total infected population roughly equal to the expected size of the GCC of the infection graph (see analysis above), while populations that took countermeasures which limited the total number of infected at or around the point of reaching herd immunity had significantly fewer total infected, ending the disease much closer to the point of our herd-immunity model.
    }
    \vskip-1em
\label{fig:pop_sim_tot_inf}
\end{figure}

We considered several limited countermeasures which may cause the disease to end sooner after reaching herd immunity (i.e. without an after-burn effect), simulate their effect upon the end of the disease for various values of $R_0$ and compare them both to our model and to the case where no countermeasures are taken.
It should be noted that the effect of the various countermeasures depends upon the temporal aspect of the spread of the disease as well as many other scenario-specific details. Our simulations provide examples of the phenomenon at hand and should not be taken as general statements. The results of our simulations are shown in figures \ref{fig:pop_sim_temporal} and \ref{fig:pop_sim_tot_inf}.


\section{Limitations of Graph Models}
The models we used throughout this paper overlook the details of the contact graph and assume it is a complete graph.
This is clearly not a realistic case, and it is possible that the details of the graph can drastically affect the outcomes.

To showcase the importance of the details of the graph architecture we will consider two examples of graphs where the measured parameters of the infection on which we have based our model are the same, but in one case we will reach herd immunity only after $90\%$ of the population was infected, while in the other we will reach herd immunity only after $o(1)$ of the population has been infected.

These examples show the limitations of random graph-based models, and why it is important
when confronting real world pandemics (or any other information spread scenario) 
to acquire as much as possible a detailed knowledge of the 
infection graph architecture. Such details can be collected by various techniques
ranging from questionnaires to contact tracing.
A mathematical framework suitable in order to overcome these shortcomings is
that of a multi-type Galton Watson branching process \cite{toappear}.

\subsection{Large Households}

Consider the following scenario: The population is divided up into households with $10$ individuals each. Whenever an individual is infected, they will infect the rest of their household, and $Poisson(1)$ other random individual.
This case is equivalent to modeling each household as a single unit with $Poisson(10)$ edges to other households, and will reach herd immunity only after a percentage of $100(1-\frac{1}{R_0}) = 90\%$
of the households has been infected (or $99.995\%$ including the after-burn effect).

A simple method commonly used to estimate the basic reproductive number of a new disease (see e.g. \cite{R01, R02, R03, R04}) is to compute the base of the exponential graph of the number of infected over time and factor in the time delay between exposure and infectiousness. Thus, the estimation method takes the number of infected as a function of time $I(t)$ and the expected depth of the infection tree as a function of time $d(t) \approx \frac{t}{\tau}$ (where $\tau$ is the approximate infection time), and tries to make the fit $I(t) \approx c {R_0} ^ {d(t)}$, and derive $R_0$ from it.

In our special case we can easily find out what this estimation of $R_0$ will yield:
Let us denote by $I(d)$ the number of people infected (for the first time) at step $d$, of those we will denote by $H(d)$ those infected from within their household and by $R(d)$ those infected not from within their household.
That is: 
\begin{itemize}
    \item $I(d) = H(d) + R(d)$ (everybody who was infected at step $d$ was either infected from within their household or not).
    \item $H(d+1) = 9 R(d)$ (only people who weren't infected from within their household can infect the other $9$ people in their household).
    \item $R(d+1) = H(d) + R(d)$ (anyone can infect outside their household)
\end{itemize}

These equations give us a simple recursion formula of 2 values. Let $M=\begin{bmatrix}0 & 9 \\ 1 & 1\end{bmatrix}$ be the step matrix, i.e. 
\begin{equation}
\begin{bmatrix}H(d+1) \\ R(d+1)\end{bmatrix} = M \begin{bmatrix}H(d) \\ R(d)\end{bmatrix} \ .
\end{equation}
Since 
\begin{equation}
I(d) = \begin{bmatrix}1 & 1\end{bmatrix} \begin{bmatrix}H(d) \\ R(d)\end{bmatrix} = M^d \begin{bmatrix}H(0) \\ R(0)\end{bmatrix} \approx constant \cdot \max(eigenvalues(M))^d \ ,
\end{equation}
we consider the eigenvalues of $M$. The characteristic polynomial of $M$ is $\lambda^2 - \lambda - 9$ and its roots are $\frac{1 \pm \sqrt{37}}{2}$. Therefore $\lambda_{max} = \frac{1 + \sqrt{37}}{2}\approx 3.5$, and this is also the value of $R_0$ that would be obtained by such an estimate.

Furthermore, we can also study the eigenvector of the matrix $M$ that produces this eigenvalue, to know what will be the approximate ratio between $H(d)$ and $R(d)$. This shows us that $\frac{R(d)}{H(d)}\rightarrow \frac{1+\sqrt{37}}{18} \approx 0.39$. That means that 
\begin{equation}
\frac{R(d)}{I(d)} = \frac{R(d)}{R(d) + H(d)} =\frac{1+\sqrt{37}}{19 + \sqrt{37}} \approx 28\% 
\end{equation}
of the newly infectious will infect 9 members of their respective households as well as another random individual, and are thus responsible for 
\begin{equation}
\frac{10 R(d)}{10 R(d) + H(d)}\approx 80\%
\end{equation}
of the next generation of infections

Though this example is most probably more extreme than reality, it shows a case where our model would expect to reach herd immunity after only a small percentage of the population would have been infected, while in reality we would end up with $90\%$ infected ($99.995\%$ with after-burn).

\subsection{An Immune Population}

In our second example we will consider the case where $1 - \varepsilon, \varepsilon \ll 1$ fraction of the population are immune and only $\varepsilon$ are susceptible/infectious (to a much higher degree). In such a case, so long as significantly less than $\varepsilon$ of the population was infected we would not see the effects of the large immune population, causing estimates for $R_0$ and for the distribution of infectiousness and susceptibility based upon the spread of the disease to appear far worse than they really are.
Clearly the total number of infected until reaching herd-immunity / finishing the after-burn scales with $\varepsilon$ and so can be arbitrarily small.

\section{Discussion and Outlook}

The spread of the COVID-19 pandemics is characterized by high variance
of the infection and susceptible distributions.
In addition to the high degree of heterogeneity in infectiousness and
susceptibility, one expects a significant correlation between them
stemming, for instance, from the social aspect of the spread of diseases.
We studied the implications of this structure on the
condition to reach herd immunity.

We proved two claims, one for general distributions and one for the Gamma
distribution and showed that the heterogeneity and correlation have
a drastic effect on the estimate of the percentage of the population that must contract the disease before herd immunity is reached.
Under the assumption of Gamma distribution, we found that for COVID-19 
a fraction $\varepsilon\approx5\%$ of infected population suffices to reach herd immunity while a fraction $\varepsilon\approx9\%$ is needed
for COVID-2. 

We studied quantitatively the after-burn effect: After herd-immunity is achieved, it may still take a while for the disease to end and if there is a large infected population at the threshold, it may still infect many others during this decline. Using population-based simulations we showed that strict counter-measures, such as a lock-down, if initiated when herd immunity is reached, can reduce significantly the after-burn effect.
Thus, we conclude that unlike what one may be tempted to conclude, when reaching herd immunity
one should not lighten the precautions taken to slow the spread of the disease. As we showed, if no action is taken a significant percentage of the population may be infected even after reaching herd-immunity.

In the literature, 
some models aim at reaching herd immunity while others attempt
to consider the end of the disease and naturally reaching different conclusions.
We explained the difference between the models and analyzed it in the simulations.
As we showed, the herd-immunity model, despite neglecting an important aspect of the pandemic, may be more realistic in some cases, since in order to reach a strong after-burn effect, a large number of individuals must be infected when we reach herd immunity and this does not happen when strict counter-measures are employed at peaks of the disease or if a final lockdown begins just as the basic reproduction number goes below $1$.
In our simulations we considered diverse counter measures as well as not taking counter measures at all.
We also highlighted some of the limitations of these models, showing why their results should be applied very carefully.

Several recent works considered the effects of heterogeneity of infectious diseases.
In \cite{science} the authors consider a society divided into age cohorts with disease transfer rates between and within the cohorts based upon the degree of interaction between them. While the separation into subgroups of the population and their interactions adds a level of valuable information, they only consider a simple infectiousness-susceptibility distribution (which is the main reason for the distance between our herd immunity projections) and produce their results from numerical simulations.
In \cite{Herd1, herd2, herd3} herd immunity model is being modeled or simulated 
with similar conclusions to ours for the Gamma distribution.
Agent based simulations not performed in these papers
and their analysis of the after-burn effect is less detailed.

A more detailed GCC based analysis of the after-burn was performed in \cite{He}. This analysis follows the previous work of \cite{PRE_GCC} which predicts the size of the GCC of undirected graphs for arbitrary degree distribution. In particular this analysis is applied to negative binomial distributions and results in (slightly) less optimistic projections than ours. We generalized their result which assumes an undirected infection graph (and implicitly a full correlation between infectiousness and susceptibility), and 
supported it with agent based simulations.


\section*{Acknowledgements}

We would like to thank Nir Kalkstein and Aviad Rubinstein for valuable discussions on the importance of the high variance to the spread of the disease.
The work is supported in part by the Israeli Science Foundation center
of excellence, the European Research Council (ERC) under the European Union’s Horizon 2020 research and innovation program (Grant agreement No. 835152), 
 BSF 2016414  and the IBM Einstein Fellowship at the Institute for Advanced Study in Princeton.

\end{document}

%% file: preamble.tex
\usepackage[utf8]{inputenc}

\usepackage{color}
\usepackage{graphicx}
\graphicspath{ {./Images/} }
\usepackage{wrapfig}
\usepackage{amsmath, amsthm, amssymb}

\usepackage{theoremref}

\newcommand\Prob[2]{{\Pr_{#1}\left[ {#2} \right]}}
\newcommand\cProb[3]{{\Pr_{#1}\left[ \left. #3 \;\right\vert #2 \right]}}

\newcommand\Expect[2]{{\mathop{\mathbb{E}}_{#1}\left[ {#2} \right]}}

\newcommand\defeq{\stackrel{def}{=}}

\usepackage{soul}
\usepackage{xcolor}

